\documentclass[a4paper,11pt]{article}

\usepackage{multicol}
\usepackage{graphicx}
\usepackage{setspace}
\usepackage[
   pdftitle={New expression for the K-shell ionization}, 
   pdfauthor={J. P. Santos, M. Guerra, and F. Parente},         
   pdfkeywords={electrons, atoms, collisions, ionization},   
   colorlinks,
   linkcolor=blue,
   urlcolor=blue,
   citecolor=blue,
   a4paper=true,
   ]{hyperref}

\newcommand{\abstracttitle}[1]{
 \begin{center}{\Large {\bf #1}}\end{center}
}
\newcommand{\authors}[1]{
 \vspace*{-0.3cm}
 \begin{center} {\bf #1} \end{center}
 \vspace*{-0.3cm}
}
\newcommand{\addresses}[1]{
 \begin{center} {\small #1} \end{center}
}
\newcommand{\synopsis}[1]{
 \begin{center}
 \setstretch{0.75}
 \begin{minipage}[t]{16cm}
   {\footnotesize {\bf Synopsis} #1 }
 \end{minipage}
 \setstretch{1.0}
 \end{center}
}
\newcommand{\abstracttext}[1]{
 \vspace*{-0.3cm}
 \columnsep0.75cm
 \begin{multicols}{2} #1 \end{multicols}
}

\newcommand{\picturelandscape}[2]{
 \vspace*{0.5cm}
 \centerline{
  \includegraphics*[width=7.8cm,angle=#1]{#2}
 }
}
\newcommand{\capt}[2]{
 \vspace*{-0.3cm}
 \begin{center}
 \begin{minipage}[t]{7.8cm} {\small {\bf Figure~#1}.~#2} \end{minipage}
 \end{center}
 \vspace*{0.3cm}
}

\newcommand{\writeto}[1]{
 \hspace*{-2.5mm} \footnote{E-mail: \href{mailto:#1}{#1}}\hspace*{-1.5mm}
}

\begin{document}

\abstracttitle{
New expression for the K-shell ionization
}

\authors{
J. P. Santos$^{\ast}$\writeto{jps@fct.unl.pt},
M. Guerra$^{\ast }$\writeto{mguerra@fct.unl.pt},
F. Parente$^{\ast }$
}

\addresses{
$^\ast$ Centro de F\'isica At\'omica, CFA, Departamento de   F\'isica, Faculdade de Ci\^encias e Tecnologia, FCT, Universidade Nova de Lisboa, 2829-516 Caparica, Portugal}

\synopsis{
A new expression  for the total K-shell ionization cross section by electron impact based on the relativistic extension of the binary encounter Bethe (RBEB) model, valid from ionization threshold up to relativistic energies, is proposed.  The new MRBEB expression is used to calculate the K-shell ionization cross sections by electron impact for the selenium atom. Comparison with all, to our knowledge, available experimental data shows good agreement.
}

\abstracttext{

The study of several fields, such as radiation science, plasma physics, astrophysics, elemental analysis using X-ray fluorescence, Auger electron spectroscopy, electron energy loss spectroscopy, and electron probe microanalysis, requires a large and continuous amount of K-shell ionization by electron impact cross section values. 

The complete quantum-mechanical description of the ionization by electron impact of multi-electron atoms and ions is a problem with a high degree of complexity.
During the past two decades, several powerful computer-intensive  \textit{ab initio} models have  been developed  to calculate electron impact ionization cross sections, achieving remarkable agreement with experimental ionization data for targets as hydrogen, helium, and sodium. The price for this agreement is the long time required even with the most powerful computers and, more importantly, limited applicability to targets with complex valence shell structures and heavy atoms.

Considering that most of the above mentioned applications  require at least 25\% accurate cross sections for a wide range of targets and energies, and the limitations of the accurate \textit{ab initio} calculations,  many analytical formulas have been developed in the last few years to provide reasonably accurate and reliable theoretical methods at low, medium and high incident energies to calculate electron impact total ionization cross sections for a large number of neutral atoms in a simple and flexible way..

The relativistic extension of the binary encounter Bethe (RBEB) model, proposed by Kim, Santos and Parente~\cite{YKK00}, is one of the methods that meet these requirements for neutral atoms using an analytic formula that requires the target particle's binding ($B$) and kinetic energy ($U$). 
The only \textit{ad-hoc} term introduced in the RBEB model, the Burguess-Vriens scaling factor, accounts for the projectile's kinetic energy change upon entering the atomic cloud. 

We have developed a new model  for the total K-shell ionization cross section based on the RBEB model, called MRBEB, that clarifies the origin and role of the Burguess-Vriens scaling factor~\cite{MG11}. This model uses a different scaling and requires only the particle's binding energy  ($B$) of the target.

\picturelandscape{0}{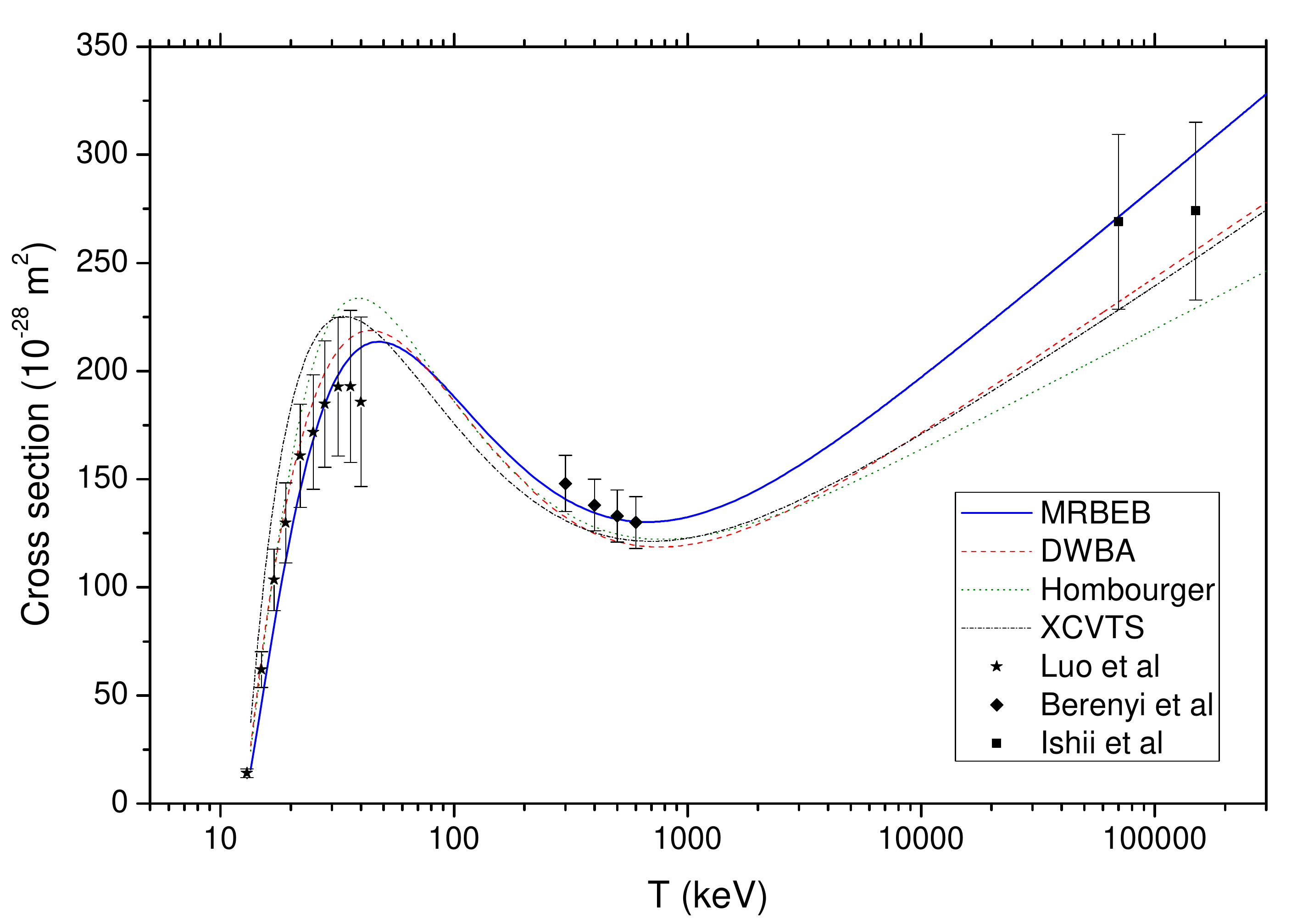} 
\capt{1}{Electron impact K-shell ionization cross sections for Se. }

In Fig. 1 we present the MRBEB cross sections for the selenium atom, as well as the theoretical calculations using the DWBA, Hombourger and XCVTS models, and the available experimental data by Luo \textit{et al}, Berenyi \textit{et al}, and Ishii \textit{et al}. We observe that the MRBEB model is theoretical approach that follows closer the experimental data.


}  

\begingroup
\small
\begin{thebibliography}{9}

\bibitem{YKK00}
Y.K. Kim, J.P. Santos, F. Parente, (2000) {\em Phys. Rev.} A {{\bf 62} 052710}
\bibitem{MG11}
M. Guerra, F. Parente, J.P. Santos, (2011) {\em}  {{\bf To be submitted} }


\end{thebibliography}
\endgroup
\end{document}